\documentclass[preprint,12pt]{elsarticle}

\usepackage{amssymb}
\usepackage{amsmath}

\usepackage[dvipsnames,table,xcdraw]{xcolor}
\usepackage{tcolorbox}
\usepackage{multirow}
\usepackage{tabularx}
\usepackage{colortbl}
\usepackage{booktabs}
\usepackage{ragged2e}
\usepackage{url}
\newcolumntype{C}{>{\Centering\arraybackslash}X}

\usepackage{parskip}
\usepackage{framed}

\usepackage[utf8]{inputenc}
\usepackage[T5]{fontenc}
\usepackage{lmodern}  
\usepackage{tipa}
\usepackage{seqsplit}
\usepackage{regexpatch}
\usepackage{xurl}         

\usepackage{graphicx}
\usepackage{multirow}
\usepackage{tabularx}
\usepackage{float}
\usepackage{booktabs}
\usepackage{nicematrix}
\usepackage{makecell}
\usepackage{adjustbox}

\usepackage{hyperref}
\usepackage{xcolor}

\definecolor{myhighlightcolor}{RGB}{255, 250, 220}
\definecolor{shadecolor}{gray}{0.9}
\newenvironment{graybox}{\begin{shaded}\small}{\end{shaded}}

\journal{Computer Speech \& Language}

\begin{document}

\begin{frontmatter}


\title{TSPC: A Two-Stage Phoneme-Centric Architecture for
Code-Switching Vietnamese-English Speech Recognition}

\author[addr1]{Tran Nguyen Anh}
 \ead{trannguyenanh280303@gmail.com}
\author[addr1]{Truong Dinh Dung}
 \ead{truongdinhdung0212@gmail.com}
\author[addr1]{Vo Van Nam}
 \ead{vvnam1812@gmail.com}
\author[addr1]{Minh N. H. Nguyen\corref{cor1}}
 \ead{nhnminh@vku.udn.vn}

\address[addr1]{The University of Danang, Vietnam - Korea University of Information and Communication Technology, Danang,Vietnam}
\cortext[cor1]{Corresponding author}

\begin{abstract}
Code-switching (CS) presents a significant challenge for
general Auto-Speech Recognition (ASR) systems. Existing methods often fail to capture the subtle phonological shifts inherent in CS scenarios. The challenge is
particularly difficult for language pairs like Vietnamese
and English, where both distinct phonological features
and the ambiguity arising from similar sound recognition are present. In this paper, we propose a novel
architecture for Vietnamese-English CS ASR, a Two-Stage Phoneme-Centric model (TSPC). TSPC adopts a phoneme-centric approach based on an extended Vietnamese phoneme set as an intermediate representation for mixed-lingual modeling, while remaining efficient under low computational-resource constraints. Experimental results demonstrate that TSPC consistently outperforms existing baselines, including PhoWhisper-base,
in Vietnamese-English CS ASR, achieving a significantly
lower word error rate of 19.06\% with reduced training
resources. Furthermore, the phonetic-based two-stage
architecture enables phoneme adaptation and language
conversion to enhance ASR performance in complex CS
Vietnamese-English ASR scenarios.
\end{abstract}




\end{frontmatter}



\section{Introduction}
\label{sec:intro}
Automatic Speech Recognition (ASR) has achieved substantial progress in recent years, significantly improving the quality of human-machine interaction. However, the seamless recognition of code-switching (CS) remains a persistent challenge, particularly in scenarios where speakers naturally alternate between languages within a conversation. Despite the increasing popularity of multilingual E2E models, ASR systems often fail to resolve fine-grained phonetic distinctions arising from cross-lingual phonological overlap.
As demonstrated in Table \ref{tab:introtab}, typical ASR models suffer significant performance degradation when applied to Vietnamese-English CS speech. The existing models frequently exhibit systematic phonetic confusion, where English lexical items are incorrectly transcribed into phonetically similar Vietnamese words, such as ``concert'' being transcribed as ``con sót''. Hence, leveraging language-specific phonological cues is challenged to distinguish cross-lingual acoustic overlaps.

\begin{table}[htb!]
\centering
\label{tab:introtab}
\resizebox{\textwidth}{!}{%
\setlength{\tabcolsep}{4pt}
\begin{NiceTabular}{p{5cm}|p{5.2cm}|p{5.2cm}|p{5.2cm}}
\toprule
\textbf{Label} & \textbf{PhoWhisper-Large}\cite{phowhisper} &\textbf{Whisper-Large} \cite{radford2023robust} & \textbf{mms-1b-all} \cite{pratap2023mms} \\
\midrule
thứ ba đó là \textit{\textbf{thinking}} hay là \textit{\textbf{feeling}}&thứ ba đó là \textit{\textbf{thìn kinh}} hay là \textit{\textbf{phiệu lìn}}h& Thứ ba đó là \textit{\textbf{thình kinh}} hay là \textit{\textbf{phiểu linh}}&th ba o la thin kinh hay la \textbf{phiu linh}\\ \addlinespace
\midrule
khi mình đi dự \textit{\textbf{concert}}&khi mình đi giữ \textbf{con sót}&khi mình đi giữ \textbf{con sót}&khi minh di d \textbf{con sot}\\ \addlinespace

\bottomrule
\end{NiceTabular}%
}
\caption{The results shown by standard models, where the English words in \textbf{bold} are incorrectly transcribed.}
\end{table}

A major limitation in current E2E paradigms is the embedding of phonological structures through high-level semantic representations. Although contextual biasing techniques, such as the bias-encoder in Deep Context \cite{pundak2018deep}, attempt to mitigate this problem by injecting domain knowledge, they may struggle with distributional overlap in large-scale corpora. Similarly, explicit Language Identification (LID) frameworks \cite{zeng2018end, lid-moe} are often bottlenecked by the data scarcity of naturalistic code-switched corpora for low-resource languages. Although recent attention-based adaptation methods \cite{chu2025adacs} offer improved generalization, they rarely account for the prosodic and tonal dimensions that define the phonetic inventory of the matrix language.

Moreover, Vietnamese is a tonal language characterized by six distinct lexical tones \cite{huu2023mispronunciation}. The interplay between English phonemes and Vietnamese tones creates a unique set of inter-lingual homophones as mentioned in \cite{duong2009mistake, anh2011l1}, such as the Vietnamese  ``\textit{lít}'' versus the English ``\textit{list}'' or ``style'' - ``xờ tai'', and ``concert'' - ``con sót'', as presented in Figure \ref{fig:cs_scenario} and Table \ref{tab:introtab}. Traditional tone-insensitive models often struggle to disambiguate such pairs, as pitch-dependent phonemic distinctions are not explicitly preserved in the acoustic-to-grapheme mapping.

\begin{figure*}[htb]
\includegraphics[width=\textwidth]{Images/cs_scenario.pdf}
\label{fig:cs_scenario}
\centering
\caption{Code-switching example in Vietnamese, where English is transformed into a Vietnamese syllable (red).} 
\end{figure*}

In this paper, we propose the TSPC model, a Two-Stage Phoneme-Centric architecture designed to bridge the gap between acoustic variability and linguistic transcription in Vietnamese-English CS. Rather than relying on a direct E2E mapping, TSPC decomposes the task into two specialized stages. First, a Speech-to-Phone (S2P) module converts acoustic input into tone-aware phoneme sequences, enabling explicit modeling of both tonal and non-tonal phonemic inventories. Second, a Phone-to-Text (P2T) module performs phoneme-to-orthography conversion, resolving lexical ambiguity through phonological constraints.

\section{Phoneme-Centric Model}
\subsection{Unified Vietnamese Phoneme Representation}

\begin{figure}[ht]
\centering
\includegraphics[width=\textwidth]{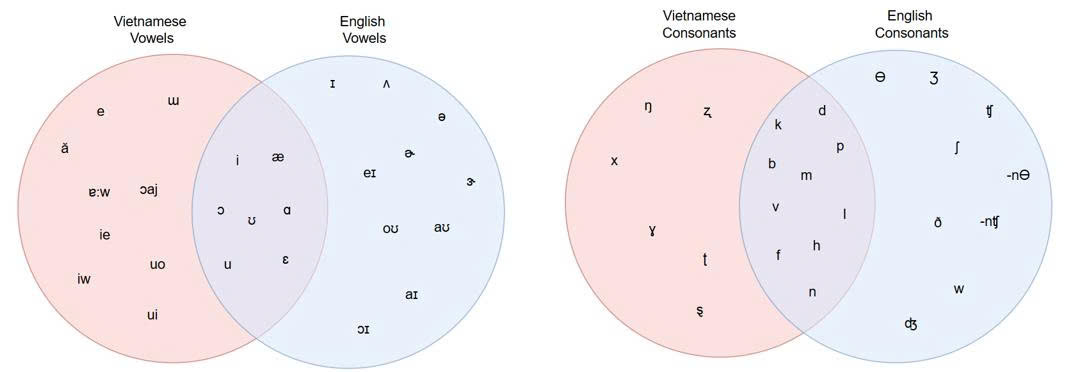}
\caption{Overlapping phonological systems of Vietnamese and English}
\label{fig:intersection}
\end{figure}

In code-switched ASR, phoneme-level representation serves as an intermediate representation bridging acoustic signals and textual output \cite{merz2022discourse, phone}. Compared to direct acoustic-to-grapheme mappings, phoneme-centric modeling enables more precise capture of linguistic structure and improves robustness across different languages and their variants.
Vietnamese and English exhibit substantial overlap in both vowel and consonant inventories, including shared phonemes such as ([p], [b], [m], [n], [i]), as presented in Figure \ref{fig:intersection}. The phonological overlap creates ambiguity in code-switched speech, where acoustically similar phonemes may correspond to lexical units from different languages. Conventional acoustic models often struggle to resolve the ambiguity due to insufficient phonological discrimination. Hence, the challenge is further complex by the tonal nature of Vietnamese, which consists of short syllables associated with six lexically contrastive tones \cite{anh2011l1}. Although English is non-tonal, Vietnamese speakers frequently adapt English pronunciations into tone-bearing syllabic forms. For example, the English diphthong \textit{``e\textipa{I}''} is commonly aligned with the Vietnamese syllable \textit{``ây''}, reflecting systematic cross-linguistic phonological adaptation.

\begin{figure*}[htb]
\includegraphics[width=\textwidth]{Images/tsne4.pdf}
\caption{t-SNE visualization of English and Vietnamese sound similarities, we use PhoWhisper-base encoder as mono-lingual model for embedding audio feature.} 
\label{fig:tsne}
\end{figure*}

\begin{table}[htb!]
\centering
\resizebox{.85\textwidth}{!}{%
\small
\begin{NiceTabular}{c c c c c c c}
\toprule
\Block{2-1}{English \\(example)} 
 & \multicolumn{3}{c}{Prefix} 
 & \multicolumn{3}{c}{Postfix} \\
\cmidrule(lr){2-4} \cmidrule(lr){5-7}
 & IPA & vi-syllable & phone 
 & IPA & vi-syllable & phone \\
\midrule
zoo   & z   & d   & z &  \textipa{u:} & u & u - 0 \\
play  & pl  & p, l & p, l &  e\textsci &  ây  & \textschwa \space - 0 iz \\
go    & g   & g   & \textgamma & \textschwa \textupsilon & âu & \textschwa \space - 0 uz \\
come  & k   & c   & k  & \textturnv m & âm & \textschwa \space - 0 mz \\
young & j   & gi  & z  & \textturnv \textipa{N} & ăng & a - 0 \textipa{N}z \\
sing  & s & s & s &  \textsci\textipa{N}   &  ing   &  i - 0 \textipa{N}z  \\
bee    &   b&  b&   b& \textipa{i:} & i   & i - 0 \\
pet    &   p&   p&   p& et & ét   & \textipa{E} - 4 tz \\
core   &   k&   c&   k& \textipa{\textopeno:}  & o   & \textopeno \space - 0 \\
foot   &   f& ph&   f& \textupsilon t & út   & u - 4 tz \\
tea & t & t & t & \textipa{i:}& i & i - 0 \\
think & $\theta$ & th & \textipa{t\super h} &  \textipa{IN}k & in & i - 0 nz \\
view & v & v & v & j\textipa{u:}& iu& i - 0 uz  \\
ship & \textesh & s & s & \textipa{I}p& íp& i - 4 pz  \\
lamp & l & l & l & \ae mp & am & \textipa{a:} - 0 mz\\
tour & t & t & t & \textupsilon\textschwa r & ua & u\textschwa \space - 0\\
\bottomrule
\end{NiceTabular}%
}
\caption{Comparative analysis of English words, an English word is separated into prefix and postfix parts, where a Vietnamese syllable (vi-syllable) is mapped with IPA, and converted to a Vietnamese phoneme (phone).}
\label{tab:ipa_syllable}
\end{table}
To analyze the cross-linguistic phonetic interactions, we conduct a comparative study using phonetic embeddings extracted from a pretrained Vietnamese ASR encoder such as PhoWhisper-Base embedding. As shown in Fig. \ref{fig:tsne}, t-SNE visualization reveals clear clustering between acoustically similar Vietnamese and English phonetic units, confirming the presence of cross-lingual phonetic similarity.

Based on the preceding observations, we construct an Unified Vietnamese phoneme representation that enables English pronunciations to be modeled within a unified Vietnamese-centric phonemic space. Instead of constructing English and Vietnamese as independently phonological systems, our approach leverages systematic phonetic similarity to align English lexical items with Vietnamese syllabic and phonemic structures.
As illustrated in Table \ref{tab:ipa_syllable}, English phonetic components are decomposed and aligned with acoustically similar Vietnamese syllables, forming an intermediate syllable-level representation. The syllabic forms are converted into tone-aware Vietnamese phoneme sequences using standardized phonetic vocabularies and predefined conversion rules. For instance, the English pronunciation ``\textit{a}'' is aligned with the Vietnamese syllable ``\textit{ây}'', which is further represented in the unified phoneme space as ``\textipa{@} - 0 iz''.
Therefore, the unified phonemic representation provides a consistent intermediate linguistic layer for modeling Vietnamese–English code-switched speech. 

\subsection{Phoneme Conversion Procedure}

\begin{figure}[ht]
\centering
\includegraphics[width=.9\textwidth]{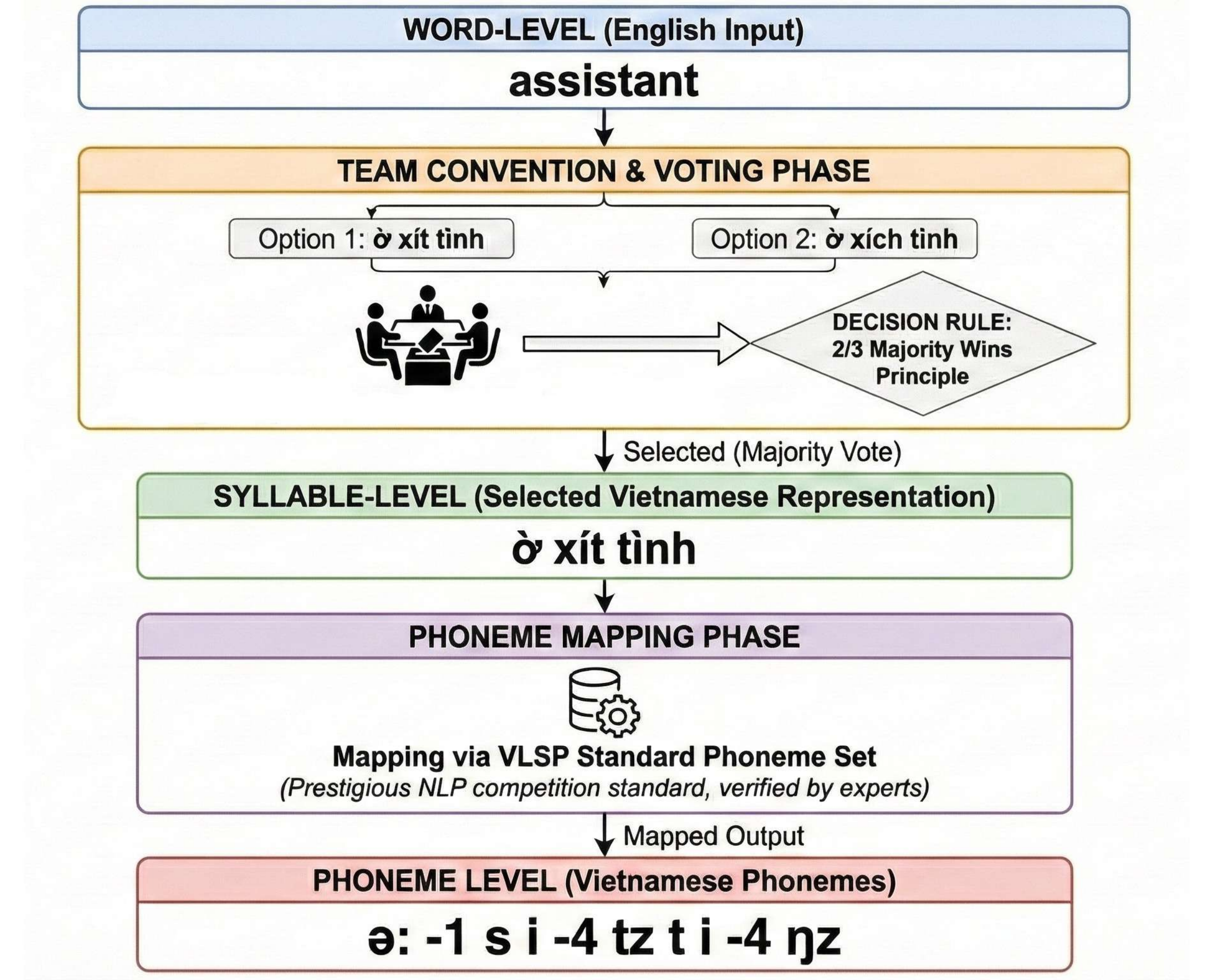}
\caption{Pipeline processing of phoneme.}
\label{fig:phone-process}
\end{figure}

The phoneme conversion procedure bridges English lexical inputs to a Vietnamese phonetic space. As shown in Fig. \ref{fig:phone-process}, English words (e.g., ``assistant'') first be designed by a Team Convention and Voting Phase, where linguistic experts propose multiple Vietnamese syllabic candidates (e.g., ``ờ xít tình'' vs. ``ờ xích tinh'') reflecting natural pronunciation. A two-thirds majority rule selects the final form to ensure consistency. The chosen syllables are then mapped to phonemes using the VLSP Standard Phoneme Set, which decomposes them into detailed phoneme sequences with tone markers, such as ``\textipa{@}: -1 s i -4 tz t i -4 $\eta z$''.
By explicitly incorporating phoneme-level structure and tone indicators (e.g.,-1,-4), the proposed conversion procedure enables English lexical items to be consistently represented within the Vietnamese phonemic space. The proposed unified representation facilitates more precise acoustic-phonetic modeling and reduces phonological ambiguity in Vietnamese–English code-switched ASR articulated within a Vietnamese linguistic framework.

\subsection{Dataset Preparation}
The P2T dataset is constructed by generating a mix of English-Vietnamese transcriptions, mapping English words to Vietnamese syllables to form initial localized code-switched text within a pre-defined vocab.
Both the mapped English-Vietnamese text and these collected Vietnamese transcriptions are then transformed into phonemes. Supporting S2T and S2P model development, a speech dataset is curated: for S2T, the code-switched text is synthesized into audio via a high-quality Text-to-Speech (TTS) service and integrated with an existing Open Vietnamese Dataset; for S2P, audio transcriptions within the S2T dataset are converted to phonetic form, the entire process of data curation is demonstrated in Figure~\ref{fig:data_curation}. 

\begin{figure*}[htb]
\includegraphics[width=\textwidth]{Images/datacuration.pdf}
\centering
\caption{Details of data curation and processing including Phone-to-Text and Speech dataset} 
\label{fig:data_curation}
\end{figure*}

Furthermore, in practice, there are various pronunciations of an English word based on individual perception, resulting in a diversity of pronunciations. To account for this diversity, each English word is matched with various syllable variants, as illustrated in Figure \ref{fig:example}, while Vietnamese transcriptions are concurrently gathered from an independent source.


    

\begin{figure}[h]
\centering
\resizebox{.9\textwidth}{!}{%
  \begin{minipage}{\columnwidth}
  \begin{graybox}
\begin{tabular}{p{0.44\columnwidth} !{\color{gray!40}\vrule width 0.5pt} p{0.44\columnwidth}}
  \parbox[t]{0.44\columnwidth}{
    \textbf{Reference:} \\
    Hôm qua tớ vừa xem cái \textbf{video} này hay lắm. \\
    (\textit{Yesterday, I watched this video, it was really cool.}) \\[6pt]

    \textbf{Input:} \\
    Hôm qua tớ vừa xem cái \textbf{vi deo} này hay lắm. \\
    (\textit{``vi deo'': Vietnamese conversion of ``video''}) \\[6pt]
    
    \textbf{Variants (word replacements):} \\
    \quad \text{video} $\rightarrow$ \text{vi đêu} (\textit{variation with different pronunciation}) \\
    \quad \text{video} $\rightarrow$ \text{vi đê ô} (\textit{different speaking of ``video''})
  }

  &

  \parbox[t]{0.44\columnwidth}{
    \textbf{Reference:} \\
    Tôi \textbf{comment} lại để xem xét sau nhé \\
    (\textit{I will leave a comment to consider later.}) \\[6pt]

    \textbf{Input:} \\
    Tôi \textbf{com men} lại để xem xét sau nhé \\
    (\textit{``com men'': Vietnamese conversion of ``comment''}) \\[6pt]

    \textbf{Variants (word replacements):} \\
    \quad \text{comment} $\rightarrow$ \text{còm men} (\textit{variation with different pronunciation}) \\
    \quad \text{comment} $\rightarrow$ \text{com mần} (\textit{different speaking of ``comment''})
  }

  \end{tabular}
  \end{graybox}
  \end{minipage}
}
\caption{Example of variants in Phone-to-Text dataset}
\label{fig:example}
\end{figure}

In our study, the corpus for training CS ASR system consists of existing Vietnamese speech datasets, including VLSP 2020 \cite{VLSP2020_ASR_eval} (92.03 hours), VietBud500 \cite{Bud500} (43.55 hours), Common Voice \cite{ardila2019common} (20.3 hours), LSVSC \cite{lsvsc} (49.17 hours), and VSV \cite{VietSpeech} (31.51 hours). In addition, we incorporated 7.32 hours of Vietnamese-English CS speech, which includes both the Capleaf \cite{capleaf_viVoice_2024} and synthetic CS data. 
For CS evaluation, a subset of 1.18 hours of CS speech was used to assess the system performance.

\subsection{Two-stage Model Development}

The proposed two-stage ASR model utilizes phonemic representation by combining two foundation models, such as Speech-to-Phone (S2P) and Phone-to-Text (P2T) that are independently pre-trained, then integrated and fine-tuned for code-switched scenarios.

\subsubsection{Speech-to-Phone}
For the S2P model, we adopt the well-established Sequence-to-Sequence (Seq2Seq) paradigm \cite{sutskever2014sequence} for speech-to-text tasks due to its efficient architecture for learning complex acoustic features and mapping variable-length input speech to a sequence of phonemes. Specifically, the S2P model employs a pre-trained encoder built on large-scale Vietnamese datasets that can capture rich acoustic feature extraction capabilities. Following the acoustic encoder, different variants of decoders are validated for generating phoneme sequences. The overall process of S2P is formulated as follows: 
\begin{align}
H &= Encoder_{no\_grad}(x_1, x_2, ..., x_n) \\
P &= Decoder(H)
\end{align}
Given an input speech sequence $x_1, x_2, \dots, x_n$, where $n$ denotes the number of frames in the mel-spectrogram, the pre-trained encoder extracts contextual acoustic representations $H$. The subscript $no\_grad$ indicates that the encoder is frozen and used solely for feature extraction. The decoder then maps hidden state $H$ to the phoneme sequence $P$, modeling the acoustic-to-phoneme transformation. Based on the comprehensive experiments 
we identify the most suitable decoder for phoneme recognition, as summarized in Table
\ref{tab:performance_results}. 

\begin{table}[hbt!]
\centering
\resizebox{\textwidth}{!}{
\begin{tabular}{lcccc}
\toprule
\multirow{2}{*}{\textbf{Encoder - Decoder}} & \multicolumn{4}{c}{\textbf{Phone Error Rate (\%) $\downarrow$}} \\ 
\cmidrule(lr){2-5}
 & \textit{LSVSC} & \textit{Vietbud\_500} & \textit{CmV} & \textit{VLSP 2020} \\
\midrule
PhoWhisper - GRU        & 1070  & 2230  & 3280  & 1970  \\
PhoWhisper - LSTM       & 720   & 1290  & 1410  & 1403  \\
Wav2vecVN - Transformer & 46.20 & 37.73 & 17.66 & 49.09 \\
\midrule
\textbf{PhoWhisper - Transformer} & \textbf{8.43} & \textbf{5.71} & \textbf{10.13} & \textbf{15.4} \\
\bottomrule
\end{tabular}
}
\caption{Vietnamese phoneme recognition results of architectural combinations.}
\label{tab:performance_results}
\end{table}

\subsubsection{Phone-to-Text}\label{sec:p2t}
Phone-to-Text (P2T) conversion is framed as a translation problem, drawing inspiration from Machine Translation (MT) \cite{stahlberg2020neural}. In this work, the phoneme sequence serves as the \textit{``source language''} and the text as the \textit{``target language''}. We choose the T5 model \cite{raffel2023exploringlimitstransferlearning} for the P2T model, which frames all NLP problems as text2text generation tasks. However, when integrated with the S2P model, prediction errors such as incorrect or spurious phonemes introduce noise into the P2T stage, degrading transcription quality, especially severely in the CS setting, where phoneme errors occur more frequently and propagate to the downstream model.
To mitigate the problem, we adopt a masking strategy inspired by prior masked modeling approaches \cite{bert}, as illustrated in Figure \ref{fig:maskingdata}. 

\begin{figure}[ht]
\centering
\includegraphics[width=\textwidth]{Images/maskingdata.pdf}
\caption{Process of constructing masking text-based samples, where an English word in code-switching case is transformed into a Vietnamese syllable before converted phoneme and randomly masked.}
\label{fig:maskingdata}
\end{figure}

We pre-train the encoder with a phoneme masking objective to learn robust contextual representations under noisy inputs, and then use it as a feature extractor for the decoder. During fine-tuning, we consider two strategies: fully freezing (fully freeze) the encoder, or freezing only the first three layers (layers freeze) while fine-tuning the remaining layers for better task adaptation. The process of building the P2T model is denoted as:

\begin{equation}
\begin{aligned}
H &= \text{Encoder}_{\theta_{\text{no\_grad}}}(X), \quad \text{where}
&\begin{cases}
\theta \in N, \quad \text{or} \\
\theta_{n} \le 3, \quad n \in N
\end{cases} \\
O &= \text{Decoder}(H)
\end{aligned}
\end{equation}

where $X, \text{ and } N$ represent the list of phonemes, and the number of layers, respectively.

\subsubsection{Two-Stage Phoneme-Centric Model}\label{sec:tspc}

\begin{figure*}[ht]
\includegraphics[width=\textwidth]{Images/tspc_model.pdf}
\centering
\caption{Two-Stage Phoneme-Centric model (TSPC), where predicted phoneme sequence represents the input for P2T model.}
\label{fig:two-stage-model}
\end{figure*}

The integration of two separate models into a unified architecture is inherently challenging. Although we apply self-supervised learning (SSL) to the P2T encoder to enhance its ability to robustly capture phonetic context, the overall integration is not straightforward. In practice, a naive combination of the two models without careful parameter tuning can even degrade translation performance, as observed in our experiments. To overcome the issue, we combine S2P and P2T modules and perform joint fine-tuning as shown in Fig. \ref{fig:two-stage-model}. 
In particular, we freeze S2P parameters during the tuning phase to ensure consistent phonetic sequence production. The P2T model is continually updated to adapt to predicted phonemes. In addition, we explore three fine-tuning strategies to better understand the integration dynamics: (1) full fine-tuning, (2) partial fine-tuning following a strategy similar to the P2T approach, and (3) fine-tune the encoder only.

\subsubsection{Training Objective}

In the setting of training objective for three models, we train the models by minimizing the cross-entropy loss: 
\begin{equation}
\mathcal{L}_{\mathrm{CE}}
=
- \sum_{t=1}^{T}
\log P\left(y_t^{\mathrm{true}} \mid y_{<t}, x\right)
\end{equation}
where $T$ is sequence length, and $\mathcal{L}_{\mathrm{CE}}$ is used for Phoneme Recognition, Masked Language Modeling, and Machine Translation tasks, where $x$ can be seen as phone or token level inputs. In the joint-finetune stage, we use token-level loss as the P2T is solely updated.

\section{Experiments}
\subsection{Training Details}
In pre-training, the S2P model used a frozen PhoWhisper-base encoder for acoustic feature extraction and trained its Transformer decoder for 15 epochs, while the P2T model was trained on phoneme–text pairs for 40 epochs, and in the MLM task, we follow the implementation similar to \cite{roberta}, with the number of duplications for each sample is $5$ and pretrained with 30 epochs. Both models share a 6-layer encoder–decoder architecture with a model dimension of $512$, using a batch size of $16$, AdamW optimizer, $lr = 1e - 4$, and linear warm-up scheduling. During joint fine-tuning, the TSPC model was trained for $20$ epochs with batch size $8$ and $lr = 3e-5$. The implementation was based on PyTorch and trained on a single NVIDIA GTX 3090 GPU.

\subsection{Experimental Results}

\renewcommand{\arraystretch}{0.8}
\begin{table*}[htb]
\centering
\resizebox{\textwidth}{!}{%
\begin{tabular}{l | l | l l l l l}
\toprule
\multirow{2}{*}{\textbf{Model}} & 
\multirow{2}{*}{\textbf{CS}} & 
\multirow{2}{*}{\makecell{\textbf{Vi} \\ \textbf{avg}}} & 
\multicolumn{4}{c}{\textbf{Vi}} \\
\cline{4-7}
& & & \rule{0pt}{2.6ex}\textit{LSVSC} & \textit{Vietbud\_500} & \textit{CMV} & \textit{VLSP 2020} \\
\midrule
facebook/mms-1b-all & 100.43 & 93.14 & 92.49 & 92.86 & 91.88 & 95.33\\
openai/whisper-base & 59.45 & 74.83 & 52.01 & 92.94 & 92.9 & 61.50 \\
openai/whisper-large-v3-turbo & 31.60 & 45.23 & 27.77 & 70.31  & 41.19 & 41.65\\
Qwen3-ASR-0.6B & 38.93 & 22.47 & 12.65 & 10.49 \textcolor{ForestGreen}{(-3.81)} & 40.97 & 25.76\\
\midrule
Wav2vec2-vn-base & 38.06 & 21.70 & 13.71 & 16.40 & 32.42 & 24.31 \\
vinai/PhoWhisper-base & 27.90 & \textbf{14.05} & \textbf{9.40} & 14.33 & \textbf{16.08} & \textbf{16.42} \\
\midrule
TSPC (baseline) & 25.35 & 18.13 & 14.90 & 11.59 & 20.93 & 25.13 \\
\quad + w/ Joint FT & 19.90 \textcolor{ForestGreen}{(-8)} & 16.47 \textcolor{Red}{(+2.42)} & 13.64 & 10.63 & 18.30 & 23.33 \\
\quad + w/ SSL P2T enc + joint FT &
\textbf{19.06} \textbf{\textcolor{ForestGreen}{(-8.84)}} & 15.87 \textcolor{Red}{(+1.82)} & 12.42 &
\textbf{9.94} \textbf{\textcolor{ForestGreen}{(-4.36)}} & 18.01 & 22.39 \\
\bottomrule
\end{tabular}%
}
\caption{WER results of methods on Code-switching (CS) and Vietnamese (Vi) test sets. Best results in \textbf{bold}. The improvement and decrease of WER are compared to PhoWhisper-base as the SOTA model and presented in green and red, respectively.}
\label{tab:combined_wer_results}
\end{table*}

\subsubsection{Code-switching speech recognition}

Table \ref{tab:combined_wer_results} demonstrates the effectiveness of TSPC in improving code-switching (CS) performance. Starting from a baseline of 25.35\%, Joint fine-tuning (Joint FT) reduces the CS error rate to 19.90\%. Incorporating the SSL P2T encoder with Joint FT further improves performance, achieving the best overall CS score of 19.06\% and outperforming all other systems. In contrast, Qwen3-ASR-0.6B \cite{Qwen3-ASR} (38.93\%), Wav2Vec2-vn-base \cite{wav2vec2_vi_2021} (38.06\%), and PhoWhisper-base (27.90\%) yield substantially higher CS error rates. Although these pretrained models remain competitive, the proposed TSPC variants—particularly when integrated with the SSL P2T encoder — deliver the most significant gains in code-switching robustness.

\subsubsection{Vietnamese speech recognition}
For the Vietnamese setting in Tab. \ref{tab:combined_wer_results}, PhoWhisper-base achieves a strong 14.05\%, outperforming Qwen3-ASR-0.6B (22.47\%) and Wav2Vec2-vn-base (21.7\%), highlighting the benefit of Vietnamese-specialization. Despite being built with more limited resources, the TSPC variants deliver highly competitive results: from a baseline of 18.13\%, Joint FT reduces overall performance on the Vietnamese test to 16.47\%, and the SSL P2T encoder + joint FT further improves to 15.87\%. On the Vietnamese subsets, the SSL P2T variant attains the best Vietbud\_500 score (9.94\%) and shows consistent gains across CMV and VLSP 2020.

\subsection{Ablation Studies}
As mentioned in the process of building P2T and TSPC models, we explore different parameter-freezing strategies to study fine-tuning effectiveness in the two-stage framework. The P2T model uses two settings (i.e., fully or partially frozen masked encoder) while TSPC evaluates three approaches during joint fine-tuning (i.e., full finetune, partially frozen, encoder tuned only), aiming to identify the most effective fine-tuning combination for the two-stage model. The results in Table \ref{tab:p2t_results} show that with a pretrained P2T encoder by the SSL approach and layer freezing during finetune with the decoder, the BLEU score of code-switching cases (CS) rises to 92.90\% and other sets also moderately improve, while fully freezing degrades performance, as the frozen pretrained encoder cannot adapt to the MT task. 

\begin{table}[htb]
\centering
\resizebox{1.\textwidth}{!}{%
\begin{tabular}{l | l | l | l l l l}
\toprule
\textbf{Model} & \textbf{Freeze strat. enc} & \textbf{CS} & \textbf{LSVSC} & \textbf{Vietbud\_500} & \textbf{CMV} & \textbf{VLSP 2020} \\
\midrule
P2T (baseline) & \textbf{--} & 92.11 & 99.50 & 99.84 & 98.33 & 96.40 \\
\midrule
\multirow{2}{*}{P2T w/ SSL enc.} & layers freeze & \textbf{92.90} \textbf{\textcolor{ForestGreen}{(+0.79)}} & \textbf{99.64} & 99.84 & \textbf{98.52} & \textbf{96.48} \\
& fully freeze & 88.75 \textcolor{Red}{(-3.36)} & 97.88 \textcolor{Red}{(-1.62)} & 99.72 & 98.23 & 95.37 \textcolor{Red}{(-1.03)} \\
\bottomrule
\end{tabular}
}
\caption{Comparison of P2T (baseline) and P2T with masked encoder (SSL enc.), the improvement and decrease are compared to the baseline (results measured by BLEU).}
\label{tab:p2t_results}
\end{table}

In the setup of the pretrained P2T model during joint-finetuning (Table \ref{tab:ablation_wer_results}) shows that joint-finetune strategies (JFT Type) strongly impact performance.
With the baseline setup (full finetune), this reduces CS to 19.25\% and improves VLSP 2020 to 22.04\% in the setup of fully freeze P2T encoder in the prior stage. More remarkably, the setting of fine-tune only encoder achieves the best CS of 17.78\% and further gains on CMV 18.00\% and VLSP 2020 22.39\%. In contrast, layers freeze yields higher CS (19.59\%–21.35\%), showing weaker adaptation.

\renewcommand{\arraystretch}{0.8}
\begin{table*}[htb]
\centering
\resizebox{\textwidth}{!}{%
\begin{tabular}{l | l | l | l | l l l l}
\toprule
\multirow{2}{*}{\textbf{Model}} &
\multirow{2}{*}{\textbf{JFT Type}} & 
\multirow{2}{*}{\textbf{Freeze strat. enc.}} &
\multirow{2}{*}{\textbf{CS}} & 
\multicolumn{4}{c}{\textbf{Vi}} \\
\cline{5-8}
 & & & & \rule{0pt}{2.6ex}\textit{LSVSC} & \textit{Vietbud\_500} & \textit{CMV} & \textit{VLSP 2020} \\
 
\midrule
\multirow{2}{*}{Pretrained P2T w/ joint FT}
& \multirow{2}{*}{full} & layers freeze (baseline) & 19.25 & 12.92 & 9.95 & 18.02 & 22.60 \\
&  & fully freeze & 20.73 \textcolor{Red}{(+1.48)} & 12.48 & 9.93 & 17.34 & \textbf{22.04} \textbf{\textcolor{ForestGreen}{(-0.56)}} \\

\midrule
\multirow{2}{*}{Pretrained P2T w/ joint FT}
& \multirow{2}{*}{layers freeze} & layers freeze   & 19.59 & 13.25 & 10.17 & 18.52 & 23.02 \\
& & fully freeze & 21.35 & 12.73 & \textbf{9.85} \textbf{\textcolor{ForestGreen}{(-0.1)}}  & 17.94 & 22.23 \\

\midrule
\multirow{2}{*}{Pretrained P2T w/ joint FT}
& \multirow{2}{*}{encoder only} & layers freeze & 17.78 & 13.21 & 10.29 & 18.58 & 23.03 \\
& & fully freeze & \textbf{19.06} \textbf{\textcolor{ForestGreen}{(-0.19)}} & 12.42 & 9.94  & 18.00 & \textbf{22.39} \textbf{\textcolor{ForestGreen}{(-0.21)}} \\

\bottomrule
\end{tabular}%
}
\caption{Comparative analysis of P2T during joint-finetuning, where Freeze strat. enc. represents the freezing strategy for the pretrained P2T encoder, and JFT Type is the joint-finetuning strategies of the TSPC model, the improvement and decrease compared to the baseline (results measured by WER).}
\label{tab:ablation_wer_results}
\end{table*}

\section{Discussion}
Our findings illustrate the importance of the intersection in the phonetic system between Vietnamese and English, particularly in low-resource code-switching settings. The intersection strategy helps mitigate hallucination when code-switching data is limited by providing more stable phonetic grounding during decoding. In addition, translating English sounds into Vietnamese syllable-based representations according to phonetic similarity, as illustrated in Figure \ref{fig:tsne}, establishes a fundamental framework for handling pronunciation variability and effectively addresses challenges posed by non-native Vietnamese speakers.

In addition, the two-stage architecture demonstrates strong potential at the phoneme level by leveraging phonetic representations as a structured intermediate space and exploiting phoneme context to generate the final target sentence. As shown in Table \ref{tab:combined_wer_results}, when compared with PhoWhisper (pretrained on 800 hours including private data), our model achieves 15.87\% on the overall Vietnamese test, only 1.82\% higher despite substantially lower resource usage, and notably outperforms Wav2VecVN (fine-tuned on 250 hours) by a margin of 5.83\%. The results highlight the adaptability and efficiency of the proposed architecture, particularly under computational constraints. Furthermore, PhoWhisper is a Vietnamese-specialized model and includes code-switching cases in its training data, achieving 27.90\% on the CS test, the TSPC variants still demonstrate competitive and superior performance, ranging from 25.35\% down to 19.06\%.

However, although the results show clear potential, several limitations remain. Firstly, the S2P model is trained on only $200$ hours of data due to limited computing resources, which is not enough to cover the wide range of Vietnamese and code-switching cases, and errors at the phoneme level directly affect the P2T model when the input phoneme sequence is wrong. Secondly, creating synthetic data is difficult in terms of maintaining diversity and good audio quality, which limits generalization. Thirdly, phonemes have their own structure, and although applying masking on the P2T encoder slightly improves performance, the method is still not sufficient for a standard Transformer to fully learn and model the structure of phonemes.

The observations open up opportunities for future work. To better capture relationships between phonemes, incorporating graph-based modeling could provide clear advantages, as graphs explicitly represent structural relations. Previous studies, such as GraphRAG \cite{graphrag} and GraphMERT \cite{graphmert}, have shown the importance of modeling symbolic and syntactic relations at the token level, which can be extended to phoneme-level representations. A Vietnamese syllable consists of multiple structural components, and when words are decomposed into phonemes, they form flexible phone sequences; therefore, explicitly modeling the structural relations among phonemes and their syntactic roles within a sentence becomes necessary, and graph-based approaches could help address this limitation. In addition, pretraining the decoder and fine-tuning it jointly with the encoder remains beneficial in low-resource settings, as shown in Table \ref{tab:combined_wer_results}. When the P2T decoder is frozen to preserve its text generation ability, the model still achieves 19.06\%, indicating that maintaining pretrained knowledge is effective even without full fine-tuning.

\section{Conclusion}
In our study, we proposed a two-stage phoneme-centric architecture designed for Vietnamese-English CS speech recognition. 
The extensive experiments demonstrate the efficiency of Vietnamese phoneme representation by enhancing the model's ability to handle the mix of Vietnamese-English languages. Notably, the proposed approach is well-suited for low-resource settings, maintaining strong performance even with limited training data and computational capacity. The proposed method emphasizes the importance of converted phonemes in advancing the speech recognition performance in code-switching and future multilingual systems.

\bibliographystyle{elsarticle-num-names} 
\bibliography{SpeechCom/main}

@inproceedings{pundak2018deep,
  title={Deep context: end-to-end contextual speech recognition},
  author={Pundak, Golan and Sainath, Tara N and Prabhavalkar, et. al.},
  booktitle={2018 IEEE spoken language technology workshop (SLT)},
  pages={418--425},
  year={2018},
  organization={IEEE}
}

@article{zeng2018end,
  title={On the end-to-end solution to mandarin-english code-switching speech recognition},
  author={Zeng, Zhiping and Khassanov, Yerbolat and Pham, Van Tung, et. al.},
  journal={arXiv preprint arXiv:1811.00241},
  year={2018}
}

@misc{lid-moe,
      title={Enhancing Code-Switching Speech Recognition with LID-Based Collaborative Mixture of Experts Model}, 
      author={Hukai Huang and Jiayan Lin et. al.},
      year={2024},
      eprint={2409.02050},
      archivePrefix={arXiv},
      primaryClass={cs.CL},
      url={https://arxiv.org/abs/2409.02050}, 
}

@inproceedings{chu2025adacs,
  title={AdaCS: Adaptive Normalization for Enhanced Code-Switching ASR},
  author={Chu, The Chuong and Pham, Vu Tuan Dat, et. al.},
  booktitle={ICASSP 2025-2025 IEEE International Conference on Acoustics, Speech and Signal Processing (ICASSP)},
  pages={1--5},
  year={2025},
  organization={IEEE}
}

@inproceedings{merz2022discourse,
  title={Discourse on ASR Measurement: Introducing the ARPOCA Assessment Tool},
  author={Merz, Megan and Scrivner, Olga},
  booktitle={Proceedings of the 60th Annual Meeting of the Association for Computational Linguistics: Student Research Workshop},
  pages={366--372},
  year={2022}
}

@article{phone,
 ISSN = {00978507, 15350665},
 URL = {http://www.jstor.org/stable/409603},
 author = {Morris Swadesh},
 journal = {Language},
 number = {2},
 pages = {117--129},
 publisher = {Linguistic Society of America},
 title = {The Phonemic Principle},
 urldate = {2025-09-02},
 volume = {10},
 year = {1934}
}

@article{anh2011l1,
  title={L1 influence on Vietnamese accented English},
  author={Anh, Ngo Phuong},
  journal={Voices},
  pages={108--125},
  year={2011}
}

@article{duong2009mistake,
  title={Mistake or Vietnamese English},
  author={Duong, Thi Nu},
  journal={VNU Journal of Foreign Studies},
  volume={25},
  number={1},
  year={2009}
}

@inproceedings{huu2023mispronunciation,
  title={Mispronunciation detection and diagnosis model for tonal language, applied to Vietnamese},
  author={Huu, Tuong Tu and Pham, Viet Thanh et. al},
  booktitle={Proc. INTERSPEECH},
  volume={2023},
  pages={1014--1018},
  year={2023}
}

@article{sutskever2014sequence,
  title={Sequence to sequence learning with neural networks},
  author={Sutskever, Ilya and Vinyals, Oriol et. al.},
  journal={Advances in neural information processing systems},
  volume={27},
  year={2014}
}

@article{stahlberg2020neural,
  title={Neural machine translation: A review},
  author={Stahlberg, Felix},
  journal={Journal of Artificial Intelligence Research},
  volume={69},
  pages={343--418},
  year={2020}
}

@misc{raffel2023exploringlimitstransferlearning,
      title={Exploring the Limits of Transfer Learning with a Unified Text-to-Text Transformer}, 
      author={Colin Raffel and Noam Shazeer et. al.},
      year={2023},
      eprint={1910.10683},
      archivePrefix={arXiv},
      primaryClass={cs.LG},
      url={https://arxiv.org/abs/1910.10683}, 
}

@misc{VLSP2020_ASR_eval,
  title        = {Automatic Speech Recognition for Vietnamese},
  howpublished = {\url{https://vlsp.org.vn/vlsp2020/eval/asr}},
  author       = {{Association for Vietnamese Language and Speech Processing}},
  year         = {2020},
  note         = {VLSP 2020 ASR evaluation campaign},
  key          = {VLSP2020\_ASR}
}

@misc{Bud500,
  author = {Anh Pham and Khanh Linh Tran et. al.},
  title = {Bud500: A Comprehensive Vietnamese ASR Dataset},
  url = {https://github.com/quocanh34/Bud500},
  year = {2024}
}

@article{ardila2019common,
  title={Common voice: A massively-multilingual speech corpus},
  author={Ardila, Rosana and Branson, Megan et.al.},
  journal={arXiv preprint arXiv:1912.06670},
  year={2019}
}

@article{lsvsc,
  title={Automatic Speech Recognition of Vietnamese for a New Large-Scale Corpus},
  author={Tran, Linh Thi Thuc and Kim, Han-Gyu, et. al.},
  journal={Electronics},
  volume={13},
  number={5},
  pages={977},
  year={2024},
  publisher={MDPI}
}

@misc{VietSpeech,
    author = {Pham Quang Nhut and Duong Pham Hoang Anh and Nguyen Vinh Tiep},
    title = {VietSpeech: Vietnamese social voice dataset},
    url = {https://github.com/NhutP/VietSpeech},
    year = {2024}
}

@misc{capleaf_viVoice_2024,
  title        = {viVoice: Vietnamese Multi-Speaker Speech Synthesis Dataset},
  author       = {{Capleaf}},
  year         = {2024},
  howpublished = {\url{https://huggingface.co/datasets/capleaf/viVoice}},
  note         = {Hugging Face dataset (gated access; CC-BY-NC-SA 4.0)},
  key          = {viVoice2024}
}

@misc{phowhisper,
      title={PhoWhisper: Automatic Speech Recognition for Vietnamese}, 
      author={Thanh-Thien Le and Linh The Nguyen et. al. },
      year={2024},
      eprint={2406.02555},
      archivePrefix={arXiv},
      primaryClass={eess.AS},
      url={https://arxiv.org/abs/2406.02555}, 
}

@misc{wav2vec2_vi_2021,
  author = {Thai Binh Nguyen},
  doi = {10.5281/zenodo.5356039},
  month = {09},
  title = {{Vietnamese end-to-end speech recognition using wav2vec 2.0}},
  url = {https://github.com/vietai/ASR},
  year = {2021}
}

@inproceedings{radford2023robust,
  title={Robust speech recognition via large-scale weak supervision},
  author={Radford, Alec and Kim, Jong Wook et. al.},
  booktitle={International conference on machine learning},
  pages={28492--28518},
  year={2023},
  organization={PMLR}
}

@article{pratap2023mms,
  title={Scaling Speech Technology to 1,000+ Languages},
  author={Vineel Pratap and Andros Tjandra et. al.},
journal={arXiv},
year={2023}
}

@misc{bert,
      title={BERT: Pre-training of Deep Bidirectional Transformers for Language Understanding}, 
      author={Jacob Devlin and Ming-Wei Chang and Kenton Lee and Kristina Toutanova},
      year={2019},
      eprint={1810.04805},
      archivePrefix={arXiv},
      primaryClass={cs.CL},
      url={https://arxiv.org/abs/1810.04805}, 
}

@misc{roberta,
      title={RoBERTa: A Robustly Optimized BERT Pretraining Approach}, 
      author={Yinhan Liu and Myle Ott and Naman Goyal and Jingfei Du and Mandar Joshi and Danqi Chen and Omer Levy and Mike Lewis and Luke Zettlemoyer and Veselin Stoyanov},
      year={2019},
      eprint={1907.11692},
      archivePrefix={arXiv},
      primaryClass={cs.CL},
      url={https://arxiv.org/abs/1907.11692}, 
}

@article{Qwen3-ASR,
  title={Qwen3-ASR Technical Report},
  author={Xian Shi and Xiong Wang and Zhifang Guo and Yongqi Wang and Pei Zhang and Xinyu Zhang and Zishan Guo and Hongkun Hao and Yu Xi and Baosong Yang and Jin Xu and Jingren Zhou and Junyang Lin},
  journal={arXiv preprint arXiv:2601.21337},
  year={2026}
}

@misc{graphrag,
      title={Retrieval-Augmented Generation with Graphs (GraphRAG)}, 
      author={Haoyu Han and Yu Wang and Harry Shomer and Kai Guo and Jiayuan Ding and Yongjia Lei and Mahantesh Halappanavar and Ryan A. Rossi and Subhabrata Mukherjee and Xianfeng Tang and Qi He and Zhigang Hua and Bo Long and Tong Zhao and Neil Shah and Amin Javari and Yinglong Xia and Jiliang Tang},
      year={2025},
      eprint={2501.00309},
      archivePrefix={arXiv},
      primaryClass={cs.IR},
      url={https://arxiv.org/abs/2501.00309}, 
}

@misc{graphmert,
      title={GraphMERT: Efficient and Scalable Distillation of Reliable Knowledge Graphs from Unstructured Data}, 
      author={Margarita Belova and Jiaxin Xiao and Shikhar Tuli and Niraj K. Jha},
      year={2025},
      eprint={2510.09580},
      archivePrefix={arXiv},
      primaryClass={cs.AI},
      url={https://arxiv.org/abs/2510.09580}, 
}

\end{document}